\documentclass[12pt]{article}
\usepackage{epsf}
\usepackage{graphicx}
\usepackage{mathbbol}
\usepackage{amsmath,amssymb}
\usepackage{hyperref}
\usepackage{color}
\newcommand{\be}{\begin{equation}}
\newcommand{\ee}{\end{equation}}
\newcommand{\beq}{\begin{eqnarray}}
\newcommand{\eeq}{\end{eqnarray}}
\newcommand{\ba}{\begin{array}}
\newcommand{\ea}{\end{array}}

\newcommand{\dist}{\displaystyle}

\newcommand{\bfp}{\mbox {\boldmath $p$}}

\date{}

\begin{document}

\title{\normalsize{\bf Collapse to the Center and Ambiguity in the Asymptotic
Behavior of the Off-Shell Scattering Amplitude in Singular Three-Body Problems}}
\author{\normalsize\bf{A.~E.~Kudryavtsev\footnote{E-mail: kudryavt@itep.ru}, A.~I.~Romanov}}
\maketitle

\vspace{-8mm}
\centerline{\small{\it Institute for Theoretical and Experimental Physics,}}
\centerline{\small{\it National Research Center Kurchatov Institute,}}
\centerline{\small{\it Bolshaya Cheremushkinskaya street 25, Moscow, 117218 Russia}}

\vspace{-5mm}
\begin{abstract}
{\bf Abstract} --- We discuss some examples of equations of the three-body problem with the oscillating asymptotics at large momentum: (i) the fixed-center approximation, (ii) the unitarized equation in the fixed-center approximation, (iii) Skornyakov--Ter-Martirosyan equation, and (iv) equations with operators used in the effective field theory, i.e., which can be expanded in power-series with positive powers of momentum. We show that in the aforementioned three-body problems the situation analogous to the falling down to the center in the two-body problem takes place --- there appears an infinite number of bound states. The energy of these states is not bounded from below. In that sense the situation is close to the falling down to the center in the two-body problem.
\end{abstract}

\vspace{3mm}
\begin{center}
1.~INTRODUCTION
\end{center}
\normalsize

The three-body integral equations formulated by L.D.~Faddeev in the early 1960s \cite{Faddeev,Faddeev2} make it possible to find the three-body observables in terms of the two-body $T$ matrices $T_{ij}(\bfp,\bfp';E)$. By using some parametrization of the off-shell $T$ matrices and by solving the Faddeev equations for three-body observables, one can usually obtain an unambiguous result i.e.\ to find bound-state energies and scattering amplitudes as well as to determine the energy dependence of the phase shifts $\delta(k)$ and scattering lengths.

However the two-body $T$ matrix is not known completely in many cases, and one has to use an incomplete $T$ matrix. As an example, we could mention the Skornyakov--Ter-Martirosyan (STM) equation \cite{STM}, where the on-shell two-body amplitude in the unitarized scattering-length approximation, $f=\displaystyle\frac{1}{a^{-1}-ik}$, is only specified. The STM equation then gives ambiguous results for the three-body scattering amplitude. However if the three-body scattering length $a_3$ is additionally specified, then the description of the three-body system becomes more informative, i.e.\ yields, for example, the phase shifts
$\delta(k)$, see, e.g., \cite{Dan}.

Recently, the STM equation is actively used within the application of the effective field theory to the three-body problem. It is worth mentioning Ref.~\cite{BHvK} where the cyclic dependence of the three-body scattering length on the cutoff parameter $\Lambda$ in the STM equation was discussed. The authors of Ref.~\cite{BHvK} proposed a scheme for a cyclic renormalization via the inclusion of an additional three-body contact interaction.

In this paper we will discuss, among other things, a manifestation of the cyclic dependence of the three-body amplitudes on the upper limit of integration $\Lambda$ in various integral equations for the three-body systems. Using some examples, we will show that the cyclic dependence of the result on $\Lambda$ can be explained by the presence of the solution of the homogeneous three-body equation, which does not decrease at infinity.

\vspace{4mm}
\begin{center}
2.~FIXED-CENTER APPROXIMATION: BASIC RESULTS
\end{center}
\vspace{1mm}

In Refs.~\cite{KRG,KGR2016} we studied in detail the first example, where we discussed the properties of the equation in the fixed-center approximation (FCA) for the system of one light and two heavy particles. In this model, the interaction of a light particle (L) of mass $m_L$ with a heavy particle (H) of mass $m_H$ is determined by a constant amplitude which is equal to the scattering length $a$. Within the FCA multiple-scattering amplitude $F^{(M)}(k)$ in the momentum space is an integral of the Green's function for the light particle in the field of two (pointlike) centers. The FCA equation for the Green's function $R(p,p')$ is exactly solvable in the class of special functions. The equation for $R(p,p')$ reads
\begin{equation}\label{eq1}
R(p,p')=\pi\ln\left(\frac{p+p'}{p-p'}\right)^2+\frac{\bar{a}}{2\pi}
        \int\limits_0^{\infty}ds\ln\left(\frac{p+s}{p-s}\right)^2R(s,p'),
\end{equation}
where $\bar a=a(1+\xi)$, $\xi=m_L/m_H$. The function $R(p,p')$ carries all features of the three-body problem and can be obtained analytically. This unique possibility of analytical study of the property of a three-body system by solving Eq.~\eqref{eq1} has attracted our attention.

In Ref.~\cite{KGR2016} it was found that in the case of attraction in the LH two-body system ($a>0$) there is a one-parameter family of solutions of Eq.~\eqref{eq1} instead of its unambiguous solution. The reason for this ambiguity is that, in addition to the solution of the nonhomogeneous equation \eqref{eq1}, there exists a nontrivial solution of the corresponding homogeneous equation (below we use the index ``Hom'' for this solution):
\begin{equation}\label{eq2}
R_{\rm Hom}(p,p')=\frac{\bar{a}}{2\pi}\int\limits_0^{\infty}ds
\ln\left(\frac{p+s}{p-s}\right)^2R_{\rm Hom}(s,p').
\end{equation}
Solution of Eq.~\eqref{eq2} has the form:
$$
R_{\rm Hom}(p,p')=4\pi\mathbb{B}\sin pa\sin p'a,
$$
where $\mathbb{B}$ is an arbitrary constant (normalization factor). Therefore, a general solution of Eq.~\eqref{eq1} is the sum of a particular solution of the nonhomogeneous equation \eqref{eq1}, $R_{\rm Inh}(p,p')$ (below marked by the index ``Inh''), and an
arbitrarily normalized solution of \eqref{eq2}, i.e.,
\begin{equation}\label{eq3}
R_{\mathbb{B}}(p,p')=R_{\rm Inh}(p,p')+4\pi\mathbb{B}\sin pa\sin p'a.
\end{equation}
Notice that the nontrivial solution of the homogeneous equation \eqref{eq2} exists only for $\bar{a}>0$. Therefore, the solution of Eq.~\eqref{eq1} for $\bar{a}<0$ is unambiguous, so that $R_{\rm Inh}(p,p')$ has a constant asymptotics, see \cite{KRG}, where we studied this case. But in the case $\bar{a}>0$, the asymptotic behavior of the particular solution of Eq.~\eqref{eq1} at $p\gg p'$ is
\begin{equation}\label{eq2_2}
R_{\rm Inh}^{\rm as}(p,p')=4\pi^2\cos pa\sin p'a.
\end{equation}
Hence the asymptotic behavior of the general solution of Eq.~\eqref{eq1} for
$a>0$ has the form
\begin{equation}\label{eq3_2}
R^{\rm as}(p,p')=4\pi^2(\cos pa+b\sin pa)\sin p'a,~~{\rm where}~~b=\mathbb{B}/\pi.
\end{equation}

However solutions of equations for three particles are not known analytically and therefore must be studied numerically. Integration in the right-hand side from $0$ to $+\infty$ is replaced by integration over the finite segment $[0,\Lambda]$. As was recently found in \cite{KRG,KGR2016}, the numerical solutions obtained in this way within the segment $[0,\Lambda]$ substantially depend on the choice of $\Lambda$. Namely, in \cite{KRG,KGR2016} we found an asymptotic cyclic dependence of the solution $R_{\Lambda}(p,p')$ on the parameter $\Lambda$ for $p\gg p'$ and $p\lesssim\Lambda$. The ambiguity of the solution of the equation for the resolvent of the three-body problem was first noticed by L.D.~Faddeev \cite{Faddeev}. This model was also discussed in detail in the book of Schmidt and Ziegelmann \cite{Schmid}.

In \cite{KGR2016}, we showed that the cyclic dependence of the solution on $\Lambda$ only reflects the ambiguity of the solution of Eq.~\eqref{eq1} in infinite limits and that for $p\lesssim \Lambda$ the solution $R_{\Lambda}(p,p')$ coincides with the solution of Eq.~\eqref{eq1} with the infinite limits of integration if the parameter $b$ is related to $\Lambda$ by
\begin{equation}\label{eq4}
\cot\,a(\Lambda_{\rm cr}^{(i)}-\Lambda)=b,
\end{equation}
where $\Lambda_\mathrm{cr}^{(i)}$, $i=1,2,...$, are ``critical values'' of the parameter $\Lambda$. At $\Lambda=\Lambda_\mathrm{cr}^{(i)}$ both the solution $R_{\Lambda}(p,p')$ and the three-body scattering length $a_3$ go to infinity. As we have shown in \cite{KGR2016}, the positions of $\Lambda_\mathrm{cr}^{(i)}$ for the FCA equation are well described by the following linear equation:
\begin{equation}\label{eq7}
\Lambda_\mathrm{cr}^{(i)}=\Lambda_\mathrm{cr}^{(1)}+(i-1)\cdot\Delta\Lambda,
\end{equation}
where $\Delta\Lambda={\rm const}$.

The solutions $R(p,p')$ and $R_{\Lambda}(p,p')$ of Eq.~\eqref{eq1} are finite for all $\Lambda \neq \Lambda_\mathrm{cr}^{(i)}$. Therefore, one should treat the upper limit of integration $\Lambda$ as a physical parameter rather than an ultraviolet cutoff. The three-body multiple-scattering amplitude $F^M$ depends unambiguously on the parameter $b$:
$$
F^M = \langle\varphi|R(p,p')|\varphi\rangle,
$$
where $\varphi(r)$ is the wave function of the HH system. Fixing $b$ in terms of the three-body scattering length $a_3$, we simultaneously fix the parameter $\Lambda$, see Eq.~\eqref{eq4}. Therefore, one should treat the parameter $\Lambda$ as a quantity which is related to physical observables --- namely, to the scattering length $a_3$. The requirement formulated in \cite{BHvK} that the result of Eq.~\eqref{eq1} should be independent of the parameter $\Lambda$ seems to be unjustified.

The interpretation of the parameter $\Lambda$ as the ultraviolet cutoff was proposed in \cite{BHvK}. This interpretation of $\Lambda$ stems from analysis of perturbation-theory series. We will discuss this point by using the FCA equation \eqref{eq1} as an example. Divergences of the perturbation theory for the FCA equation were analyzed in \cite{BE}. It was shown there that the ultraviolet perturbation-theory contributions diverge, and this divergence must be regularized in all terms of the perturbation-theory series. The degree of divergence grows with the perturbation-theory order. However, at least for $a<0$, the solution of the integral equation for the total amplitude is finite, since all divergences cancel. So we see that, for $a<0$, the sum of the contributions of multiple scattering is finite, and does not require the introduction of an ultraviolet cutoff. The case of $a>0$ contains the same perturbation-theory diagrams, with their mutual cancellation. The ambiguity in the result for the FCA amplitude for $a>0$ is explained by the fact that perturbation-theory diagrams are summed into a fixed-sign series, in contrast to what we have for $a<0$. Therefore, in summation divergences appear at the cutoff $\Lambda=\Lambda_\mathrm{cr}$. These infinities should be interpreted as physical infinities in the three-body scattering amplitude $a_3$, i.e.\ as a signal of the appearance of a new level in the three-body system. Thus, using Eq.~\eqref{eq1} for the respective Green's function as an example, we show that perturbative divergences in integral equations cancel, and that the remaining divergences at $\Lambda_\mathrm{cr}$ are quite physical. Introducing a cutoff in the equation for the multiple-scattering amplitude, we therefore accomplish a technical procedure that makes it possible to solve the problem numerically, so there is no need for interpreting $\Lambda$ as an ultraviolet cutoff. It follows that the requirement formulated in \cite{BHvK} that the result should be independent of $\Lambda$ in the limit of $\Lambda\to\infty$ is redundant. If the result converges as a function of $\Lambda$, then it is the result for the function $R(p,p')$. However, it may turn out that at $\Lambda\to\infty$ the result is ambiguous, and the amplitude $a_3$ should additionally be defined.

We note that for $a<0$, i.e.\ in the case of repulsion, the solutions of Eq.~\eqref{eq1} are unambiguous, with no cyclic dependence of the result on $\Lambda$ in that case. Simultaneously, there are no nontrivial solutions of the homogeneous equation \eqref{eq2}.

Thus, the exactly solvable example of the FCA equation demonstrates the existence of two possible realizations of solutions for the three-body problem: (i) in one of them, the solution approaches asymptotically a constant in the limit of $\Lambda\to\infty$, the respective homogeneous equation has only a trivial (zero) solution; (ii) in the other, the introduction of $\Lambda$ leads to a solution that has a cyclic rather than a decreasing character, and the homogeneous equation has a nontrivial solution.

Below, we will see that, for different three-body equations, there is similar relationship between the cyclic character in $\Lambda$ and the existence of a nontrivial solution of the respective homogeneous integral equation.

\vspace{4mm}
\begin{center}
3. OTHER EQUATIONS WITH SOLUTIONS HAVING OSCILLATING ASYMPTOTICS AT $p\to\infty$
\end{center}
\vspace{1mm}

{\it 3.1. Unitarized Equation in the Fixed-Center Approximation (UFCA Equation)}
\vspace{3mm}

We consider again the system of a light (L) particle and two heavy (H) particles, but now the LH interaction has the form
\begin{equation}\label{eq8}
f_\mathrm{LH}=\frac{1}{a^{-1}-ik}.
\end{equation}
This problem was studied earlier in \cite{Amado}. In our further consideration, we will restrict ourselves to the case of $a>0$. Numerical calculations reveal that the use of the unitary amplitude \eqref{eq8}, i.e.\ UFCA, improves substantially the accuracy of calculations for the three-body system comparing to FCA calculations (at $\xi=m_\mathrm{L}/m_\mathrm{H}\ll 1$).

The UFCA equation for the Green's function has the form [cf.\ Eq.~\eqref{eq1}]
\begin{equation}\label{eq9}
R^U(p,p')=\pi\ln(p,p')+\frac{1+\xi}{2\pi}
\int\limits_0^{+\infty}\frac{ds}{a^{-1}+g(s)}\ln(p,s)R^U(s,p'),
\end{equation}
where $g(s)=\sqrt{2m_\mathrm{LH}\varepsilon+s^2\displaystyle\frac{2\xi+\xi^2}{(1+\xi)^2}}$, $m_\mathrm{LH}=\displaystyle\frac{m_\mathrm{L}M_\mathrm{H}}{m_\mathrm{L}+M_\mathrm{H}}$, $\varepsilon$ is the binding energy of the HH system, and $\ln(p,s)\equiv\ln\left(\displaystyle\frac{p+s}{p-s}\right)^2$.

The multiple-scattering amplitude $F_U^{(M)}$ in terms of $R^U(p,p')$ can be expressed as
\begin{equation}\label{eq10}
\begin{split}
&F_U^{(M)}=2\frac{(1+\xi)^2}{1+\frac{1}{2}\xi}\iint dpdp'pp'\varphi(p)\frac{1}{a^{-1}+g(p)}\,
\frac{(4\pi)^2}{(2\pi)^6}~\times\\&~~~~~~~\times R^U(p,p')\frac{1}{a^{-1}+g(p')}\,\varphi(p').
\end{split}
\end{equation}
In contrast to the FCA case considered above, solutions of the UFCA equation \eqref{eq9} are not known analytically. However, the asymptotic behavior for the homogeneous equation corresponding to Eq.~\eqref{eq9} can be found in the region $p\gg a^{-1}$, $p\gg p'$. Indeed, the corresponding homogeneous equation in this limit ($\xi\ll 1$) looks like
\begin{equation}\label{eq11}
R^U_\mathrm{Hom}(p,p')=\frac{1}{2\pi}\int\limits_0^{+\infty}\frac{ds}{s\sqrt{2\xi}}
\ln\left(\frac{p+s}{p-s}\right)^2 R^U_\mathrm{Hom}(s,p').
\end{equation}
Using the formula
$$
\frac{2\pi}{\alpha}\frac{\cos(\theta\alpha)}{\sin(\pi\alpha)}=
\int\limits_0^{+\infty}\ln(1+2x\cos\theta+x^2)x^{\alpha-1}dx,
$$
which is known from tables of Mellin transforms, see, for example Tables of Integral Transforms by the Bateman Manuscript Project \cite{Beit}, we find that the function $R^U(p)\sim p^{\alpha}$ (as well as $p^{-\alpha}$) is a solution of Eq. \eqref{eq11} under the condition that $\alpha$ is a root of the equation
\begin{equation}\label{eq12}
\tan\frac{\alpha\pi}{2}=\alpha\sqrt{2\xi}.
\end{equation}
The first (pure imaginary) solutions of Eq.~\eqref{eq12} appear at
$$
\alpha\approx\pm \frac{i}{\sqrt{2\xi}}.
$$
Notice that there are two independent solutions of the homogeneous equation. One of them coincides with the asymptotic solution of the nonhomogeneous equation; the other is an independent solution of the nonhomogeneous equation. Taking into account that the function $R(p,p')$ is symmetric with respect to $p\leftrightarrow p'$, we construct the following linear combination of solutions:
$$
R^U_\mathrm{Hom}(p,p')\propto\cos\left(\frac{1}{\sqrt{2\xi}}\ln\frac{p}{p'}\right).
$$

Thus, there is also a nontrivial solution of the corresponding homogeneous equation for Eq.~\eqref{eq9} in UFCA, so the solution of Eq.~\eqref{eq9} depends on the normalization factor $\tilde{b}$ of the solution $R^U_\mathrm{Hom}$ of the homogeneous equation, i.e.\ depends on one parameter, as in the case of the FCA approximation.

The asymptotic behavior of the general solution of Eq.~\eqref{eq9} is
$$
R^U(p,p')\approx\sin\left(\frac{1}{\sqrt{2\xi}}\ln\frac{p}{p'}\right)+
\tilde{b}\cos\left(\frac{1}{\sqrt{2\xi}}\ln\frac{p}{p'}\right).
$$
Thus, the solution of Eq.~\eqref{eq9} depends on the normalization factor $\tilde{b}$ and is ambiguous. On the other hand, numerical solving Eq.~\eqref{eq9} within a finitelength segment $[0,\Lambda]$ show a cyclic dependence on the parameter $\Lambda$. Analyzing these numerical solutions, one can also find that, for $p\lesssim \Lambda$ and $p\gg p'$, the solution reads
$$
R^U_{\Lambda}(p,p')\propto \sin\left(\sqrt{2\xi}\ln\frac{p}{\Lambda}+\varphi\right),
$$
where the phase shift $\varphi$ does not depend on $\Lambda$.

Thus, solutions of the UFCA equation also depend cyclically on the parameter $\Lambda$ and therefore do not have a specific limit at $\Lambda\to\infty$ (however, the dependence on $\Lambda$ is softer here, since the results now depend on $\cos\ln p/\Lambda$ rather than on $\cos p/\Lambda$ as in the FCA case). At the same time, the general solution of the UFCA equation in an infinite interval depends on the normalization factor of the solution $R^U_\mathrm{Hom}(p,p')$ of the homogeneous equation.

\vspace{5mm}
{\it 3.2. Skornyakov--Ter-Martirosyan (STM) Equation}
\vspace{4mm}

The properties of solutions of the STM equation \cite{STM} and its generalizations were discussed in \cite{BHvK}. In particular, it was found that the solutions of the STM equation in the finite-length segment $[0,\Lambda]$ and of the homogeneous STM equation over an infinite interval both have an asymptotic behavior which is not decreasing for $p\gg k$. Thus, the same regularity manifests itself for the STM equation --- a cyclic dependence of the solution in the range of $[0,\Lambda]$ on $\Lambda$ and the presence of a nontrivial solution of the corresponding homogeneous equation. If the solution has a cyclic dependence on $\Lambda$, this means that there exists a nontrivial solution of the corresponding homogeneous equation with infinite limits of integration.

\vspace{10mm}
{\it 3.3. Equation for the Amplitude and Its Solution upon Expanding Two-Body Amplitudes in Series in Positive Powers of $k^2$ (FCA Plus EFT)}
\vspace{4mm}

In effective field theories, an expansion of operators in series in positive powers of momentum is used instead of the ordinary effective-range expansion for the scattering amplitude:
$$f=a+ck^2+...\;,$$
see, e.g., \cite{BE}. We now discuss the situation, using $c=2\xi a^3$, for example. The parameter $c$ has the dimensionality of $m^{-3}$. The parameter $a$ is used here on the basis of dimensional considerations in order to avoid new dimensional quantities. Instead of the FCA equation, we then obtain the following equation:
\begin{equation}\label{eqq1}
R(p,p')=\pi\ln(p,p')+\frac{a}{2\pi}\int\limits_0^{\Lambda}\left(1+2\xi a^2s^2\right)
\ln(p,s)R(s,p')ds.
\end{equation}
At small values of $\Lambda$, the first term in the parenthetical expression in the integrand dominates, and the equation under consideration reduces to the FCA equation. Further, as $\Lambda$ grows, the contribution of the second term increases, and the solutions of the two equations become substantially different.

The behavior of solutions of these equations is in qualitative agreement with the behavior of solutions of the FCA equation. Namely, there are critical values of the cutoff parameter $\Lambda$, at which the solution goes to infinity. The behavior of the solution of Eq.~\eqref{eqq1} as a function of $\Lambda$ is cyclic, but the corresponding period is smaller than in the FCA case. In order to confirm this statement, we give several critical values of $\Lambda$ for Eq.~\eqref{eqq1} at $a=0.005\text{ МэВ}^{-1}$ in the table (see also figure). One can see that now the positions of the critical values depend on their number nonlinearly.

\begin{center}{\bf Table}\end{center}
\begin{center}\begin{tabular}{|c|c|c|}
  \hline
  FCA & UFCA & EFT \\
  \hline\hline
  445 & 490 & 335 \\
  \hline
  1145 & 1260 & 643 \\
  \hline
  1845 & 2035 & 1118 \\
  \hline
  2545 & 2810 & 1435 \\
  \hline
\end{tabular}\end{center}
\begin{figure}[h!]
\centering
\includegraphics[scale=1.0]{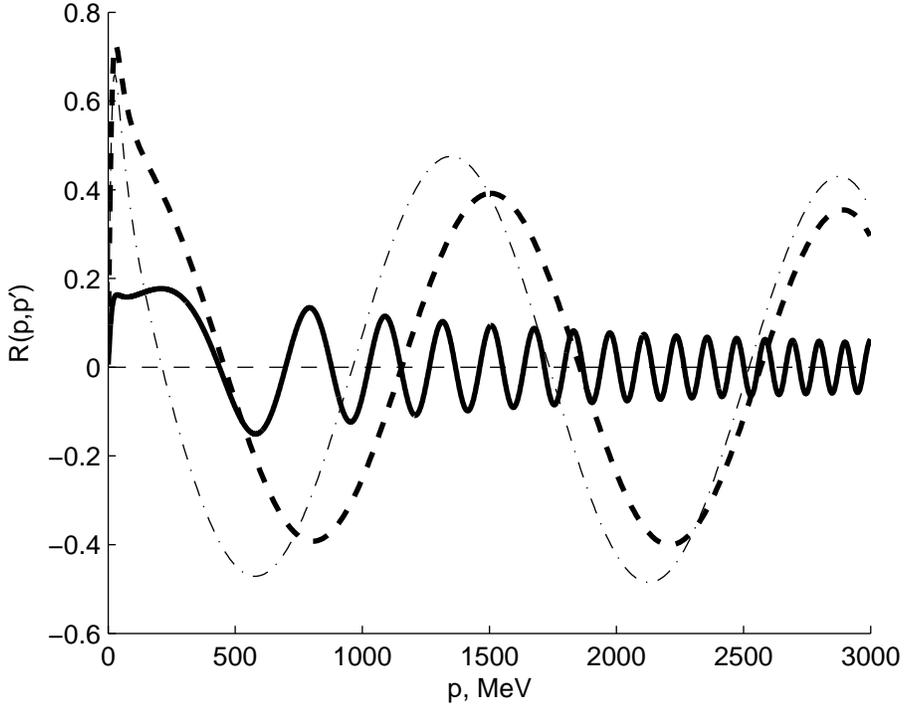}
\caption{Solution of the equation for the function $R(p,p')$ in the FCA (thick dashed curve), UFCA (dash-dotted curve), and FCA$\,+\,$EFT (solid curve) models for $\Lambda=3000$ MeV.}
\label{Fig.1}
\end{figure}

\vspace{4mm}
\begin{center}
4. AMBIGUITY OF THE SCATTERING AMPLITUDES IN THE TWO-BODY SYSTEMS WITH SINGULAR POTENTIALS
\end{center}
\vspace{1mm}

The aforementioned ambiguity of solutions of the three-body problem is not a quite novel property. Such ambiguities can be found in the two-body systems with singular potentials, in which the phenomenon of falling down to the center may occur. As an example, we consider the attractive potential $V(r)=-\beta/r^2$. The bound states and the problem of falling down to the center is discussed in sufficient detail in traditional textbooks on quantum mechanics, see, e.g., \cite{LL,MM}. The applicability of quantum mechanics to the problem of bound states in singular potentials was also discussed in \cite{Case}.

What can we say about phase shifts in the singular potential $V=-\beta/r^2$? For the radial function $R(r)$ in the $s$-wave we have:
\begin{equation}\label{eq_14}
\frac{d^2R}{dr^2}+\frac{2}{r}\frac{dR}{dr}+\frac{\gamma}{r^2}R+k^2R=0,
\end{equation}
where $\gamma=2m\beta/\hbar^2$ and $k^2=2mE$, $E>0$. At short distances, the term $k^2R$ in Eq.~\eqref{eq_14} can be neglected. We then seek the solution in the form $R\sim r^s$. This power-law dependence is a solution if $s$ is a root of the
equation
\begin{equation}\label{eq14_1}
s^2+s+\gamma=0,
\end{equation}
so that
$$
s_{1,2}=-\frac{1}{2}\pm\sqrt{\frac{1}{4}-\gamma}.
$$
In order to interpret the solutions, it is convenient to replace the singular potential at short distances $r<r_0$ by a regular potential using a boundary condition at $r=r_0$ via specifying the logarithmic derivative $\zeta$ of the function $\chi=rR$ at $r=r_0$:
$$\zeta=\frac{\theta}{r_0},$$
where $\theta$ is an arbitrary constant. For $r>r_0$, the function $\chi$ looks like
\begin{equation}\label{eq_15}
\chi=Ar^{s_1+1}+Br^{s_2+1}.
\end{equation}
From the matching condition, we obtain a relation between the ratio $B/A$ and the
parameter $\theta$ in the form
\begin{equation}\label{eq_16}
\frac{B}{A}=\left(\frac{\theta-(s_1+1)}{s_2+1-\theta}\right)x_0^{s_1-s_2},
\end{equation}
where $x_0=r_0/\bar{R}$ is the dimensionless distance and $\bar{R}\sim \sqrt{\gamma}/k$. For $\gamma<1/4$, it follows from Eq.~\eqref{eq_16} that, in the limit $r_0\to 0$, we have $B/A\to 0$ for any $\theta$. It follows that the phase shift is determined unambiguously for $\gamma<1/4$. At $\gamma>1/4$ from Eq.~\eqref{eq_16} we find that
\begin{equation}\label{eq_17}
\frac{B}{A}=\exp\left(-2i\varphi+2i\sqrt{\gamma-\frac{1}{4}}\,\ln\frac{r_0}{\bar{R}}\right),
\end{equation}
where $\varphi=\arctan\left(\sqrt{\gamma\!-\displaystyle\frac{1}{4}}\,/(\theta-\displaystyle\frac{1}{2})\right)$. From Eq.~\eqref{eq_17} it follows that for $r_0\to 0$, the ratio $B/A$ is ambiguous.

Let us now consider the case of large $r$, $r>\bar{R}$. Matching the solution \eqref{eq_15} with the solution of the free equation, $\chi=C\sin(kr+\delta)$, in the region $r>\bar{R}$, we obtain an equation for the phase shift $\delta(k)$:
\begin{equation}\label{eq_18}\dist
\arctan\frac{\sqrt{\gamma-\frac{1}{4}}}{\frac{1}{2}-\lambda(k)}=\sqrt{\gamma-\frac{1}{4}}
\,\ln\frac{r_0}{\bar{R}}+\arctan\frac{\sqrt{\gamma-\frac{1}{4}}}{\frac{1}{2}-\theta},
\end{equation}
where $\lambda(k)=\sqrt{\gamma}\cot(\sqrt{\gamma}+\delta(k))$. From Eq.~\eqref{eq_18} it follows that at any value of $\theta$ the phase shift $\delta(k)$ is not defined in the limit of $r_0\to 0$.

\vspace{4mm}
\begin{center}
5.~SINGULAR POTENTIALS AND LIPPMANN--SCHWINGER EQUATION FOR THE TWO-BODY SYSTEMS
\end{center}
\vspace{1mm}

Consider the problem of scattering in the potential $U(r)=-\displaystyle\frac{\beta}{r^2}$ in the momentum representation. It is convenient to use the Lippmann--Schwinger equation for the scattering amplitude $F(\vec{k},\vec{q})$, which has the form
\begin{equation}\label{eqq19}
F(\vec{k},\vec{q})=-U(\vec{q}-\vec{k})-\frac{2m}{\hbar^2}\int\frac{U(\vec{q}-\vec{s})
F(\vec{k},\vec{s})}{s^2-k^2-i0}\frac{d\vec{s}}{(2\pi)^3},
\end{equation}
where $U(\vec{s})=\int e^{i\vec{s}\vec{r}}U(\vec{r})d\vec{r}$ and
$F(\vec{k},\vec{s})=\displaystyle\frac{2\pi\hbar^2}{m}f(\vec{n},\vec{n}')$. For the $s$-wave component of the invariant amplitude, $F_s(k,s)=\displaystyle\frac{1}{4\pi}\int F(\vec{k},\vec{q})d\Omega_{\vec{q}}$, from Eq.~\eqref{eqq19} at $k=0$ we obtain the following equation for the scattering length:
\begin{equation}\label{eqq20}
F_s(0,q)=\frac{2\pi^2\beta}{q}+\gamma\int\limits_0^qF_s(0,t)\frac{dt}{t}+
\frac{\gamma}{q}\int\limits_q^{+\infty}F_s(0,t)dt.
\end{equation}
Because of the inclusion of a potential, the asymptotic behavior of the function $F_s(0,q)$ at large $q$ is determined from a solution of the homogeneous equation. We first find this solution. We have
\begin{equation}\label{eqq21}
F^\mathrm{Hom}_s(0,q)=\gamma\int\limits_0^qF^\mathrm{Hom}_s(0,t)\frac{dt}{t}+
\frac{\gamma}{q}\int\limits_q^{+\infty}F^\mathrm{Hom}_s(0,t)dt.
\end{equation}
After the substitution $F^\mathrm{Hom}_s(0,q)\propto q^s$, we obtain a condition for $s$, which coincides with \eqref{eq14_1}. Therefore, the general solution of Eq.~\eqref{eqq21} is
$$F^{Hom}_s\!=Aq^{s_1}\!+Bq^{s_2}\stackrel{q\to+\infty}{\longrightarrow}\left\{\begin{aligned}&\frac{\rm const}{q^{s}}
~~~~{\rm for}~\gamma<1/4,\;s=|\min\{s_1,s_2\}|\\&\frac{A}{\sqrt{q}}e^{i\sqrt{\gamma-\frac{1}{4}}\,\ln q}
+\frac{B}{\sqrt{q}}e^{-i\sqrt{\gamma-\frac{1}{4}}\,\ln q}~~{\rm for}~\gamma\geqslant 1/4.\end{aligned}\right.$$
Thus, for $\gamma>1/4$ the solution yields an indeterminate result for $B/A$ in the limit of $q\to+\infty$. There are two real-valued solutions: $\cos \left(\sqrt{\gamma-\frac{1}{4}}\ln q\right)$ and $\sin \left(\sqrt{\gamma-\frac{1}{4}}\ln q\right)$. Suppose that the second one is an asymptotic part of the solution of the nonhomogeneous equation \eqref{eqq21}. The general solution then reads
$$
A(\sin\varphi+b\cos\varphi),~{\rm where}~\varphi=\sqrt{\gamma-\frac{1}{4}}\ln q.
$$
Here, $b$ is an arbitrary constant. Because of its arbitrariness that the solution of the nonhomogeneous equation behaves ambiguously at infinity.

Consider the solution of Eq. (\ref{eqq21}) upon the introduction of the cutoff $\Lambda$ instead of integration to infinity. Substituting $F_s\propto q^s$, we obtain
$$
1=-\frac{\gamma}{s(s+1)}+\frac{\gamma}{s}\Lambda^s.
$$
After the substitution $F_s=Aq^{s_1}+Bq^{s_2}$, we obtain the following constraint on the constants $A$ and $B$:
$$
\gamma\left(A\frac{\Lambda^{s_1}}{s_1}+B\frac{\Lambda^{s_2}}{s_2}\right)=0.
$$
This yields the only solution for any value of $\Lambda$. The same solution determines the asymptotic behavior of the solution of the nonhomogeneous equation with a cutoff $\Lambda$. The solution depends unambiguously on $\Lambda$, but it does not have a specific limit for $\Lambda\to+\infty$.

\vspace{4mm}
\begin{center} 6.~CONCLUSION\end{center}
\vspace{1mm}

Singular quantum-mechanical potentials of the type $-\beta/r^n$ with $n\geqslant 2$ have been discussed quite rarely. In the present paper, we have considered some properties of the quantum-mechanical scattering problem in the potential $U(r)=-\beta/r^2$, see Sections 4 and 5. For such singular potentials, the scattering length is not determined unambiguously, i.e.\ it is necessary additionally  to choose, e.g., some scattering-length value. Thus, we see that, in the two-body sector, singular potentials seem to be exotic, and their discussion is likely to discredit their use.

Our discussion, however, demonstrates that in the three-body sector resulting equations for a number of problems look similar to the effective two-body scattering problem in a singular potential. Indeed, we have seen that, in the three-body sector, it is also necessary to introduce additionally a three-body scattering length, and that the spectrum of the problem is unbounded from below. The situation is analogous to that in the two-body system with a singular potential. Thus, we have shown that an exotic situation with the presence of singular potentials and falling down to the center in the two-body sector naturally arises as an effective theory in the three-body problem with a not exotic two-body interaction.

In conclusion, we note that, in the three-body problem, two-body potentials were treated as contact (zero-range) potentials. In physical applications, there is always a characteristic range of forces, $r_0\neq0$. Therefore, there arise various scenarios that depend on the ratio of $r_0$ and the scattering length. This may be illustrated also in the momentum representation. There are two characteristic parameters: the cutoff parameter $\Lambda$ and the inverse force range $r_0^{-1}$. Instead of $r_0^{-1}$, one can introduce a characteristic momentum above which the particle wave function falls down fast. If $a\gg r_0$ or $\beta\gg\Lambda$, then the solution of the equation for the function $R(p,p')$ is reaching regime of the asymptotic oscillations with the result that the aforementioned ambiguity in describing the system appears. Otherwise, the function $R(p,p')$ decreases due to the influence of the wave function long before the system reaches the argument $\Lambda$, so that the ambiguity does not appear. Thus, the characteristic values of $a/r_0$ enable to determine qualitatively the behavior of the system.

\vspace{5mm}
\centerline{ACKNOWLEDGMENTS}
\vspace{2mm}

We are grateful to Dr.~V.A.~Gani and Prof.~K.G.~Boreskov for stimulating and enlightening discussions.
This work was supported in part by the Russian Foundation for Basis Research
(project no. 16-02-00767).

\newpage


\end{document}